\documentclass[11pt]{article}
\usepackage{amsmath}
\usepackage{graphicx}
\usepackage{natbib}
\usepackage{mathrsfs}
\usepackage{bm, amsthm, amssymb}
\usepackage{multirow, booktabs}
\usepackage{url}
\usepackage{geometry}
\usepackage{setspace}
\usepackage{pdfpages}
\allowdisplaybreaks[4]
\newtheorem{theorem}{Theorem}

\newtheorem{assumption}{Assumption}

\newtheorem{remark}{Remark}

\DeclareMathOperator{\TE}{TE}
\DeclareMathOperator{\OTE}{OE}
\DeclareMathOperator{\IDE}{IDE}
\DeclareMathOperator{\CDE}{CDE}
\DeclareMathOperator{\CIE}{CIE}
\DeclareMathOperator{\IIE}{IIE}
\DeclareMathOperator{\CI}{CI}
\DeclareMathOperator*{\argmax}{arg\,max}
\def\independenT#1#2{\mathrel{\rlap{$#1#2$}\mkern2mu{#1#2}}}
\newcommand\independent{\protect\mathpalette{\protect\independenT}{\perp}}

\begin{document}

\title{\bf Randomized interventional effects in semicompeting risks, with application to a hematopoietic cell transplantation study}
\author{Yuhao Deng$^1$, Rui Wang$^2$, Tao Zhang$^3$ and Xiang Zhan$^4$ \\
{\small 1 Fred Hutchinson Cancer Center ~~ 2 University of Washington}\\
{\small 3 Soochow University ~~ 4 Southeast University}
}
\date{}
    
\maketitle

\begin{abstract}
In clinical studies, the risk of the primary (terminal) event may be modified by intermediate events, resulting in semicompeting risks. To study the treatment effect on the terminal event mediated by the intermediate event, researchers wish to decompose the total effect into direct and indirect effects. In this article, we extend the randomized interventional approach to time-to-event outcomes, where both intermediate and terminal events are subject to right censoring. We envision a random draw for the intermediate event process from a reference distribution, either marginally over time-varying confounders or conditionally given the observed history. We present the identification formula for interventional effects. We also discuss some variants of the identification assumptions. We estimate the treatment effects using nonparametric maximum likelihood estimation and propose a sensitivity analysis that incorporates a latent frailty. As an illustration, we study the effect of matched unrelated donor versus haploidentical donor on death mediated by relapse in a hematopoietic cell transplantation study with graft-versus-host disease (GVHD) as the time-varying confounder. We find that matched unrelated donor transplantation is preferable in terms of survival rates under the use of post-transplant PTCy GVHD prophylaxis for lymphoma patients.
\end{abstract}

\noindent%
\newpage

\section{Introduction: Mediation analysis for semicompeting risks}

In clinical trials, the primary endpoints are often the time to some event, such as disease onset or death. Researchers would like to study the treatment effect on the cumulative risk of this primary event. However, intermediate events may occur before the primary event in some individuals and mediate the risk of the primary event. The one-sided exclusion of intermediate events by the terminal event results in semicompeting risks \cite{fine2001semi}. From the counting process view, at each time point, the counting process of the intermediate event acts as a mediator between the treatment and the terminal event \cite{huang2021causal}. The total effect, evaluated by the difference in the potential cumulative incidences on the primary (terminal) event, consists of direct and indirect effects mediated by the intermediate event. 

Our study is motivated by a hematopoietic transplantation study for lymphoma patients, where relapse is an intermediate event before death, the primary endpoint \cite{mussetti2023haploidentical}. Different transplant modalities are associated with different event risks. For example, matched related donor transplantation from siblings typically exhibits a higher relapse rate but a lower transplant-related mortality, while haploidentical transplantation from family exhibits a lower relapse rate but a higher transplant-related mortality. Since relapse is associated with an increased risk of relapse-related mortality, the direct effect estimated from simple survival models that ignore intermediate events may be biased. Another issue in our data is that there is a time-varying confounder, graft-versus-host disease (GVHD), which can modify the hazards of relapse and mortality. GVHD mainly results from mismatches in human leukocyte antigen loci between the donor and recipient. A higher level of GVHD is usually associated with lower relapse risk and higher transplant-related mortality. Understanding the direct effect on death, i.e., transplant-related mortality, and the indirect effect on death, i.e., relapse-related mortality, can inform the selection of transplant modalities and medical interventions in future transplantation.

Causal mediation analysis has been extensively studied for fixed-time measurements with a single mediator and a single outcome. The most typical approach to mediation analysis is the natural effects framework. The total effect is decomposed into a natural direct effect and a natural indirect effect. Identifying natural effects requires sequential ignorability, i.e., no post-treatment confounding between the mediator and the outcome \cite{imai2010identification}. Dealing with time-varying confounders poses conceptual challenges in longitudinal studies. Natural effects can sometimes be interpreted as path-specific effects under some modified sequential ignorability assumptions in the presence of time-varying confounders \cite{mittinty2020longitudinal, miles2020semiparametric, tanner2022Methods, Kormaksson2024Dynamic}. The natural effects framework has been extended to time-to-event data, where potential outcomes are defined on the counting processes of events \cite{huang2021causal, vanderweele2011causal, deng2024direct}. Identification of natural effects for time-to-event endpoints requires sequential ignorability in a continuous-time manner, which excludes baseline frailty and post-treatment time-varying confounders that influence the hazards of intermediate and terminal events. Within the semicompeting risks framework, an alternative approach to studying treatment effects is via principal stratification, which either relies on strong cross-world assumptions, such as principal ignorability, or imposes model restrictions \cite{xu2022bayesian, nevo2022causal, gao2023defining, yu2025exploring}.

To target clinical interests directly, the separable effects framework has been proposed as an interventionist approach, assuming that treatment has two components and that each component exerts an effect on only one of the two events: the mediator (intermediate event) or the outcome (terminal event) \cite{robins2022interventionist}. The treatment components can be assigned different values in a hypothetical world or in a future experiment. The separable effects framework is a powerful device to deal with longitudinal or time-to-event measurements with time-varying confounders. The total effect is decomposed into a separable direct effect and a separable indirect effect. Identifying separable effects requires dismissible treatment components, which essentially excludes unmeasured confounding between the mediator and the outcome \cite{stensrud2022separable, breum2024estimation, yu2025Separable}. The validity of this identification assumption essentially requires that the causal pathways from each treatment component to each set of time-varying confounders are isolated, referred to as partial isolation \cite{stensrud2021generalized}. Although separable effects may align with clinical interests, this partial isolation condition cannot be verified in the presence of treatment-induced confounding in real-world trials using observed data when subject knowledge is absent, which limits the identification and interpretation of separable effects.

To relax restrictions on time-varying confounders, the randomized (stochastic) interventional effects framework was proposed as a randomized analog to controlled effects \cite{vanderweele2014effect, vansteelandt2017interventional, vansteelandt2019Mediation, diaz2020causal, diaz2023efficient}. The ideal scenario for studying mediation effects is to hold the mediator constant at a specific value. Directly controlling the mediator is unrealistic, but we can envision a hypothetical scenario in which the risk of the mediator process is fully controlled. The randomized interventional effects framework extends the controlled effects by introducing a randomized intervention into the mediator process. In a hypothetical world, we can vary the treatment condition while holding the distribution from which we draw the mediator, yielding the interventional direct effect. Similarly, we can vary the distribution from which we draw the mediator while holding the treatment condition, yielding the interventional indirect effect. Such a framework provides a direct solution for defining the treatment effect on a single event by controlling for competing risks. Specifically, Vanderweele and Tchetgen Tchetgen \cite{vanderweele2017mediation} considered random draws conditional on baseline covariates while marginalizing over time-varying confounders, whereas Zheng and van der Laan \cite{zheng2017longitudinal} and Lin et al. \cite{lin2017interventional, lin2017mediation} considered the version conditional on all history. This framework can accommodate path-specific effects with multiple intermediate variables \cite{vanderweele2017mediation, lin2017interventional, tai2023causal}, but has not been extended to semicompeting risks data with time-varying confounders \cite{valeri2023multistate}. 

Interventional effects are closely related to the hypothetical strategy described in the ICH E9 (R1) addendum to evaluate the treatment effect in clinical trials \cite{ICH19}. In our motivating example, estimating the direct effect of an alternative transplant modality calls for controlling the risk of relapse. This introduces a hypothetical scenario in which the relapse is a stochastic process following a specific intensity. Since GVHD is a treatment-induced confounding, natural effects are not well-defined, and separable effects are not identifiable. In the context of time-to-event outcomes, causal mediation analysis is challenging for the following reasons. First, the measurements for all variables are not defined after the terminal event. Each individual may have a different follow-up time. Second, the intermediate event counting process is monotonically increasing, so the draw of the counting process must respect this restriction. Envisioning external draws unconditionally is infeasible because the external process may contradict the terminal event process. Third, the mediation $g$-formula designed for longitudinal studies is inappropriate for continuous time \cite{lin2017mediation}. In our motivating data, GVHD, relapse, and death are monitored on a continuous time scale. Although the time-varying confounder process can be coarsened longitudinally in practice, discretizing follow-up time into periods can lose information and be computationally intensive.

In this article, we formalize the randomized interventional effects for semicompeting risks data. We first introduce the interventional effects in the presence of both baseline and time-varying confounders. We present the assumptions for identifying the counterfactual cumulative incidence function of the terminal event. The interventional direct and indirect effects are defined by contrasting counterfactual cumulative incidence functions. When time-varying confounders are present, the direct and indirect effects sum to an overall effect, which is generally not identical to the total effect. We also discuss some variants that make different assumptions about the interventional distribution. If the intermediate event process is drawn from the hazard conditional on time-varying confounders, then the direct and indirect effects of this version sum to the total effect. We illustrate the estimation procedure using likelihood-based methods. Under appropriate model specification, we propose a sensitivity analysis to account for latent confounding. We contribute to the literature on causal mediation analysis for time-to-event outcomes by considering identification and estimation on the continuous time scale. By appropriately modeling time-varying confounders, our method can accommodate longitudinal confounders as a special case. The likelihood-based approach allows estimation of multiple versions of the estimands using plug-in estimators. Finally, through a hematopoietic cell transplantation study, we estimate the direct and indirect effects of transplant modalities on death mediated by relapse using post-transplant cyclophosphamide (PTCy)-based GVHD prophylaxis for lymphoma patients.

\section{Interventional effects with baseline and time-varying confounders}

\subsection{Potential outcomes}

Suppose we have a binary treatment $Z \in \{0,1\}$. We define potential outcomes in terms of the counting processes of intermediate and terminal events \cite{huang2021causal}. Let $t^*$ be the end of the study. The occurrence of the intermediate event $d\tilde{N}_1(t;z)$ in the period $[t,t+dt)$ is a potential outcome of the treatment assignment $z$. The potential time to the intermediate event $T_1(z)$ is the time when the intermediate event potential counting process $\tilde{N}_1(t;z) = \int_0^t d\tilde{N}_1(s;z)$ jumps from 0 to 1. 
The occurrence of the terminal event may depend on the status of the intermediate event, so the occurrence of the terminal event $d\tilde{N}_2(t;z,\tilde{\bm{n}}_1(t))$ in the period $[t,t+dt)$ is a potential outcome of the treatment assignment $z$ and intermediate event process $\tilde{\bm{n}}_1(t)$ up to time $t$. The terminal event potential counting process $\tilde{N}_2(t;z,\tilde{\bm{n}}_1(t)) = \int_0^t d\tilde{N}_2(s;z,\tilde{\bm{n}}_1(s))$. Here, we use bold letters to represent the entire history of a counting process. Under the treatment assignment $z$, the natural terminal event counting process $\tilde{N}_2(t;z) = \tilde{N}_2(t;z,\tilde{\bm{N}}_1(t;z))$ and the potential time to the terminal event $T_2(z)$ is the time when $\tilde{N}_2(\cdot;z)$ jumps from 0 to 1. If $\tilde{n}_1(t)$ and $\tilde{N}_2(t;z,\tilde{\bm{n}}_1(t))$ jump at the same time, we assume that the intermediate event occurs first. We define the potential cumulative incidence function (CIF) of the terminal event under treatment $z$ as $F(t;z) = P\{\tilde{N}_2(t;z)=1\}$. 

\begin{figure}
\centering
\includegraphics[width=0.5\textwidth]{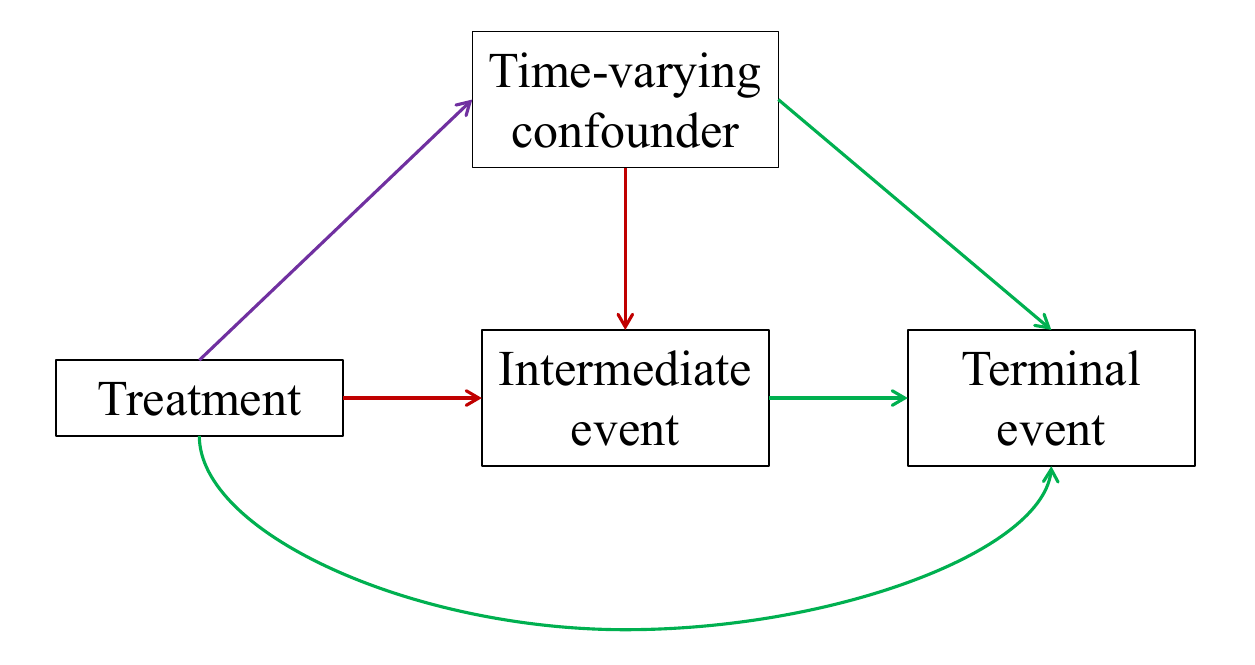}
\caption{A graph illustrating the events. Red lines indicate the paths to deliver treatment effect on the intermediate event, and green lines indicate the paths to deliver treatment effect on the terminal event. Baseline confounders are omitted for simplicity.} \label{fig:graph}
\end{figure}

In follow-up visits, time-varying confounders $L(t)$ may be collected in addition to baseline covariates $X$. Let $\bm{L}(t)$ be the history of time-varying confounders up to time $t$. Figure \ref{fig:graph} illustrates the structure of treatment, time-varying confounder, intermediate event, and terminal event. The observed follow-up time to the intermediate event $T_1$ is the minimum time of the intermediate event, the terminal event, and the censoring. The observed follow-up time to the terminal event $T_2$ is the minimum time of the terminal event and the censoring. Let $\Delta_1$ and $\Delta_2$ be the indicator of the intermediate event and terminal event, respectively, where it equals 1 if the event is observed. We need to make causal assumptions to identify the potential CIF. Let $\tilde{N}_c(t;z)$ be the potential counting process of censoring whose potential time is $C(z)$. In particular, the potential censoring process can depend on covariates $X$, time-varying confounders $\bm{L}(t)$, and the historical intermediate event process.

\begin{assumption}[Standard assumptions for identifiability] \label{asp2}
(1) Ignorability: $Z \independent (d\tilde{N}_1(t;z), d\tilde{N}_2(t;z)) \mid \tilde{\bm{N}}_1(t;z),\tilde{N}_2(t;z)=0,X,\bm{L}(t)$.
(2) Random censoring: $\tilde{N}_c(t;z) \independent (d\tilde{N}_1(t;z),d\tilde{N}_2(t;z)) \mid \tilde{\bm{N}}_1(t^-;z), \tilde{N}_2(t^-;z)=0, X, \bm{L}(t), Z=z$.
(3) Positivity: $P(Z \mid \bm{L}(t),X)>0$, and $P\{\tilde{N}_1(t;z)=n_1, \tilde{N}_2(t;z)=0 \mid \bm{L}(t),X\}>0$ implies $P\{\tilde{N}_1(t;z)=n_1, \tilde{N}_2(t;z)=0, \tilde{N}_c(t;z)=0, Z=z \mid \bm{L}(t),X\}>0$.
(4) Consistency: $T_1 = \min\{T_1(Z),T_2(Z),C(Z)\}$, $T_2 = \min\{T_2(Z),C(Z)\}$, $\Delta_1 = I\{T_1(Z) \leq T_2(Z), T_1(Z) \leq C(Z)\}$, $\Delta_2 = I\{T_2(Z) \leq C(Z)\}$.
\end{assumption}

Let 
\begin{align*}
d\Lambda_*(t;z,x,\bm{l}(t)) &= P\{d\tilde{N}_1(t;z)=1 \mid \tilde{N}_1(t^-;z)=\tilde{N}_2(t^-;z,\tilde{\bm{N}}_1(t^-;z))=0, X=x, \bm{L}(t)=\bm{l}(t)\}, \\
d\Lambda_0(t;z,x,\bm{l}(t)) &= P\{d\tilde{N}_2(t;z,\tilde{\bm{N}}_1(t;z))=1 \mid \tilde{N}_1(t;z)=\tilde{N}_2(t^-;z,\tilde{\bm{N}}_1(t^-;z))=0, X=x, \bm{L}(t)=\bm{l}(t)\}, \\
d\Lambda_1(t;z,r,x,\bm{l}(t)) &= P\{d\tilde{N}_2(t;z,\tilde{\bm{N}}_1(t;z))=1 \mid d\tilde{N}_1(r;z)=1, \tilde{N}_2(t^-;z,\tilde{\bm{N}}_1(t^-))=0, X=x, \bm{L}(t)=\bm{l}(t)\}
\end{align*}
be the hazard of the intermediate event, cause-specific hazards of the terminal event at time $t$ given the baseline confounders $x$ and a history of time-varying confounders $\bm{l}(t)$ under treatment assignment $z$ (and given that the intermediate event occurs at $r<t$ if applicable). We assume there are no ties; otherwise, if the intermediate event and the terminal event occur at the same time, we assume the intermediate event occurs right before the terminal event. Under Assumption \ref{asp2}, the hazards are identifiable,
\begin{align*}
d\Lambda_*(t;z,x,\bm{l}(t)) &= P\{t\leq T_1<t+dt, \Delta_1=1 \mid T_1 \geq t, T_2\geq t, Z=z, X=x, \bm{L}(t)=\bm{l}(t)\}, \\
d\Lambda_0(t;z,x,\bm{l}(t)) &= P\{t\leq T_2<t+dt, \Delta_2=1 \mid T_1 \geq t, T_2\geq t, Z=z, X=x, \bm{L}(t)=\bm{l}(t)\}, \\
d\Lambda_1(t;z,r,x,\bm{l}(t)) &= P\{t\leq T_2<t+dt, \Delta_2=1 \mid r\leq T_1<r+dt, T_2\geq t, Z=z, X=x, \bm{L}(t)=\bm{l}(t)\}.
\end{align*}
The proof is provided in Supplementary Material A.3. Let $N_1(t) = \Delta_1 I\{T_1 \leq t\}$ and $N_2(t) = \Delta_2 I\{T_2 \leq t\}$ be the observed counting processes of the intermediate event and the terminal event, respectively. Let $Y_1(t) = I\{T_1 \geq t\}$ and $Y_2(t) = I\{T_2 \geq t\}$ be the observed at-risk processes of the intermediate event and the terminal event, respectively. The hazards can also be written as
\begin{align*}
d\Lambda_*(t;z,x,\bm{l}(t)) &= P\{dN_1(t)=1 \mid Y_1(t)=Y_2(t)=1, Z=z, X=x, \bm{L}(t)=\bm{l}(t)\}, \\
d\Lambda_0(t;z,x,\bm{l}(t)) &= P\{dN_2(t)=1 \mid Y_1(t)=Y_2(t)=1, Z=z, X=x, \bm{L}(t)=\bm{l}(t)\}, \\
d\Lambda_1(t;z,r,x,\bm{l}(t)) &= P\{dN_2(t)=1 \mid dN_1(r)=1, Y_2(t)=1, Z=z, X=x, \bm{L}(t)=\bm{l}(t)\}.
\end{align*}
We write the cumulative hazards
\[
\Lambda_*(t;z,x,\bm{l}(t)) = \int_0^t d\Lambda_*(s;z,x,\bm{l}(s)), \ \Lambda_0(t;z,x,\bm{l}(t)) = \int_0^t d\Lambda_0(s;z,x,\bm{l}(s)), \ \Lambda_1(t;z,r,x,\bm{l}(t)) = \int_r^{r\vee t} d\Lambda_1(s;z,r,x,\bm{l}(s)),
\]
where $r\vee t = \max\{r,t\}$.

\subsection{Randomized intervention}

On the graph shown in Figure \ref{fig:graph}, there are two paths to deliver the treatment effect on the intermediate event: directly from the treatment $Z \to \tilde{N}_1(\cdot)$ or through the time-varying confounder $L(\cdot) \to \tilde{N}_1(\cdot)$. There are three paths to deliver the treatment effect on the terminal event: directly from treatment $Z \to \tilde{N}_2(\cdot)$, through the intermediate event $\tilde{N}_1(\cdot) \to \tilde{N}_2(\cdot)$, or through the time-varying confounder $L(\cdot) \to \tilde{N}_2(\cdot)$. The challenge in separating direct and indirect effects on the terminal event lies in the path from treatment to the time-varying confounder $Z \to L(\cdot)$. The treatment effect exerted through this path can contribute to either the effect on the intermediate event $Z \to L(\cdot) \to \tilde{N}_1(\cdot)$ or the effect on the terminal event $Z \to L(\cdot) \to \tilde{N}_2(\cdot)$. Suppose we are interested in a setting that allows the direct effect but controls the intermediate event. On the one hand, to leave the treatment effect via the time-varying confounder to the terminal event as natural, we should not intervene in the time-varying confounder. On the other hand, to block the treatment effect via the time-varying confounder to the intermediate event, we should intervene in the time-varying confounder. Regarding this contradiction, we need to envision a hypothetical time-varying confounder process, i.e., a recanting twin of $L(\cdot)$, and the following intermediate event process \cite{vo2026recanting}. This motivates a random draw of the intermediate event from a reference distribution that accounts for the time-varying confounder and the intermediate event process.

Suppose that we envision a random draw of the intermediate event process $G(\cdot)$ in survivors. This random draw is independent of the potential counting process of the terminal event in survivors, but can depend on historical confounders,
\begin{equation} \label{exog}
\bm{G}(t) \independent d\tilde{N}_2(t;z,\tilde{\bm{n}}_1(t)) \mid \tilde{N}_2(t^-;z,\tilde{\bm{n}}_1(t^-))=0, X, \bm{L}(t).
\end{equation}
The intervention in the intermediate event process changes the data-generating mechanism. Under treatment assignment $z$, the counting process of the intermediate event $\tilde{\bm{N}}_1(t;z)$ is replaced with an external version $\tilde{\bm{G}}(t)$. We assume that the terminal event mechanism is invariant in the hypothetical world, as that in the realized experiment under treatment assignment $z$. That is, the draw of the intermediate event process should not change the hazard of the terminal event if the drawn intermediate event process coincides with the observed one. We formalize this assumption as sequential ignorability.

\begin{assumption}[Sequential ignorability with time-varying confounders] \label{ie:haz2_t}
$\tilde{\bm{N}}_1(t;z) \independent d\tilde{N}_2(t;z,\tilde{\bm{n}}_1(t)) \mid \tilde{N}_2(t^-;z,\tilde{\bm{n}}_1(t^-))=0, X, \bm{L}(t)$.
\end{assumption}

Sequential ignorability requires no unmeasured confounding between $\tilde{\bm{N}}_1(t;z)$ and $\tilde{N}_2(t;z,\tilde{\bm{n}}_1(t))$, so that intervening in $\tilde{N}_1(t;z)$ does not change the potential counting process of the terminal event $\tilde{N}_2(t;z,\tilde{\bm{n}}_1(t))$. Together with Equation \eqref{exog}, sequential ignorability implies
\begin{equation}
\begin{aligned}
&\quad~ P\{d\tilde{N}_2(t;z,\tilde{\bm{n}}_1(t))=1 \mid \tilde{N}_2(t^-;z,\tilde{\bm{n}}_1(t^-))=0, \bm{G}(t)=\tilde{\bm{n}}_1(t), X=x, \bm{L}(t)=\bm{l}(t)\} \\
&= P\{d\tilde{N}_2(t;z,\tilde{\bm{n}}_1(t))=1 \mid \tilde{N}_2(t^-;z,\tilde{\bm{n}}_1(t^-))=0, \tilde{\bm{N}}_1(t;z)=\tilde{\bm{n}}_1(t), X=x, \bm{L}(t)=\bm{l}(t)\}.
\end{aligned}
\end{equation}
Sequential ignorability has two implications. First, if an intermediate event has not been drawn yet, then $P\{d\tilde{N}_2(t;z,\tilde{\bm{n}}_1(t))=1 \mid \tilde{N}_2(t^-;z,\tilde{\bm{n}}_1(t^-))=0, G(t)=n_1(t)=0, X=x, \bm{L}(t)=\bm{l}(t)\} = d\Lambda_0(t;z,x,\bm{l}(t))$. Second, if an intermediate event has already been drawn, then $P\{d\tilde{N}_2(t;z,\tilde{\bm{n}}_1(t))=1 \mid \tilde{N}_2(t^-;z,\tilde{\bm{n}}_1(t^-))=0, dG(r)=d\tilde{n}_1(r)=1, X=x, \bm{L}(t)=\bm{l}(t)\} = d\Lambda_1(t;z,r,x,\bm{l}(t))$.

Let $q_{0}(l(t) \mid z,x,\bm{l}(t^-))$ be the intensity of time-varying confounders at time $t$ conditional on the baseline confounders $x$ and a history of time-varying confounders $\bm{l}(t^-)$ before $t$ given that there are no intermediate events under treatment assignment $z$. Let $q_{1}(l(t) \mid z,r,x,\bm{l}(t^-))$ be the intensity of time-varying confounders at time $t$ conditional on the baseline confounders $x$ and a history of time-varying confounders $\bm{l}(t^-)$ before $t$ with the intermediate event occurring at $r<t$ under treatment assignment $z$. These two intensities are assumed to be identical to the observed intensities in treatment group $Z=z$. In the hypothetical world, the replacement of the intermediate event process by the external draw should neither change the intensity of time-varying confounders. In other words, the time-varying confounders $\bm{L}(t)$ should include all confounders between $\tilde{\bm{N}}_1(t^-;z)$ and $L(t)$.

To make the random draw interpretable, we may assume that the distribution of $G(\cdot)$ among survivors is identical to that of the intermediate event process among survivors under treatment assignment $z_1$, denoted by $G(\cdot;z_1)$. Counterfactually, we set the treatment at $z_2$ and let the time-varying confounders and terminal event process vary naturally, while the intermediate event process is replaced with $G(\cdot;z_1)$ as long as the individual does not experience the terminal event. In this way, the path operating the treatment effect on the intermediate event is controlled. When $z_2$ varies from 0 to 1 while keeping $z_1$ unchanged, the risk change of the terminal event represents the effect delivered directly to the terminal event and naturally through the time-varying confounders without passing through the intermediate event. The key to distinguishing this direct effect is to rule out the indirect effect on $G(\cdot)$ through time-varying confounders, i.e., the hazard of $G(t;z_1)$ should not depend on $\bm{L}(t)$. By integrating $\bm{L}(t)$ out, we define the hazard of the intermediate event conditional only on baseline covariates,
\begin{equation}
\begin{aligned}
d\Lambda_*(t;z_1,x) &:= P\{d\tilde{N}_1(t;z_1)=1 \mid \tilde{N}_2(t^-;z_1,\tilde{\bm{N}}_1(t^-;z_1))=0, \tilde{N}_1(t^-;z_1)=0, X=x\} \\
&= \int_{\mathcal{L}(t)} \prod_{0<s\leq t} q_0(l(s)\mid z_1, x, \bm{l}(s^-)) d\Lambda_{*}(t;z_1,x,\bm{l}(t))d\bm{l}(t),
\end{aligned}
\end{equation}
where $\mathcal{L}(t)$ is the support of $\bm{L}(t)$. The following assumption defines the hazard of the randomized intervention.

\begin{assumption}[Random draw] \label{ie:haz_t}
In the world with treatment assignment $z_2$ and the drawn intermediate event counting process $\bm{G}(t^-z_1)$, the hazard of $G(t;z_1)$ is identical to the natural hazard of $\tilde{N}_1(t;z_1)$ unconditional on $\bm{L}(t)$,
\begin{equation}
\begin{aligned}
d\Lambda_*(t;z_1,z_2,x,\bm{l}(t)) &:= P\{dG(t;z_1)=1 \mid \tilde{N}_2(t^-;z_2,\bm{G}(t^-;z_1))=G(t^-;z_1)=0, X=x, \bm{L}(t)=\bm{l}(t)\} \\
&= d\Lambda_*(t;z_1,x).
\end{aligned}
\end{equation}
\end{assumption}

\begin{figure}
\centering
\includegraphics[width=0.9\textwidth]{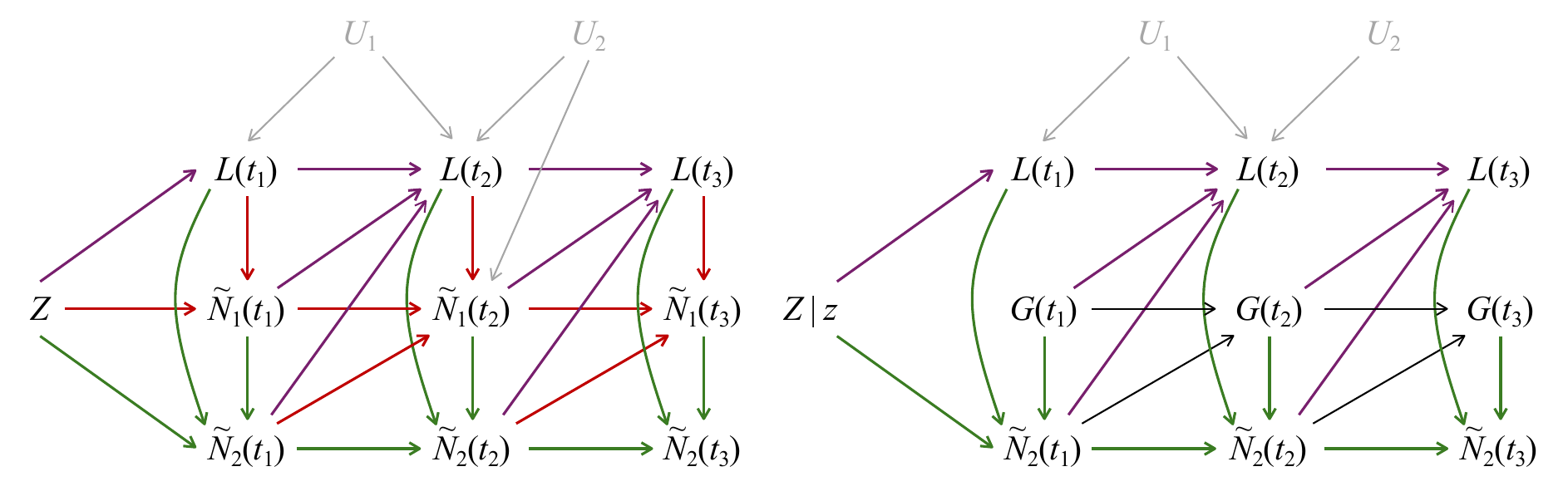}
\caption{Directed acyclic graphs (DAGs) for the counting processes. In the left figure, the counting processes are at the natural level. In the right figure, the counting process of the intermediate event is replaced by the draw $G(\cdot)$. The black arrows to $G(\cdot)$ represent deterministic relationships, in that the draw of $G(t)$ is only meaningful if $G(t^-)=\tilde{N}_2(t^-)=0$. Colored arrows represent that the treatment affects the hazards (intensities). The gray $U_1$ is an unobserved confounder within the $L(\cdot)$ process, and the gray $U_2$ is an unobserved confounder between $L(\cdot)$ and $\tilde{N}_1(\cdot)$. Baseline confounders are omitted for simplicity.} \label{fig:dag}
\end{figure}

Figure \ref{fig:dag} shows the directed acyclic graphs (DAGs) for the counting processes. In the left figure, the counting processes are at the natural level. For simplicity, we omit the arrows from $Z$ and baseline confounders $X$ to all post-treatment variables. Assumption \ref{ie:haz2_t} means that there is no unmeasured confounding between $\tilde{N}_1(\cdot)$ and $\tilde{N}_2(\cdot)$. In the right figure, the intermediate event process $\tilde{N}_1(\cdot)$ is replaced by the externally drawn process $G(\cdot)$. The unconditional hazard of $G(\cdot)$ is determined by Assumption \ref{ie:haz_t}. Since $G(\cdot)$ is an external variable, all arrows to $G(\cdot)$ shall be blocked, including the arrow from unmeasured confounder $U_2$ to $G(t_2)$. However, due to semicompeting risks, the draw $G(t)$ is only valid if $G(t^-)=\tilde{N}_2(t^-)=0$, and we use black arrows to represent this deterministic relationship. The interventional indirect effect evaluates the effect on the terminal event by manipulating the intermediate event. In Supplementary Material B, we compare the interventional effects with natural effects and separable effects.

In the hypothetical world, let
$F(t;z_1,z_2) = P\{\tilde{N}_2(t;z_2;\bm{G}(t;z_1))=1\}$
be the counterfactual CIF of the terminal event. The identifiability of the counterfactual CIF is given in the following theorem.

\begin{theorem} \label{thm:id}
Under Assumptions \ref{asp2}, \ref{ie:haz2_t} and \ref{ie:haz_t}, the counterfactual CIF of the terminal event conditional on baseline covariates $X=x$ is nonparametrically identified as
\begin{small}
\begin{equation}
\begin{aligned}
F(t;z_1,z_2,x) &= \int_0^t \int_{\mathcal{L}(s)}\prod_{0<r\leq s} q_0(l(r) \mid z_2,x,\bm{l}(r^-)) \exp\{-\Lambda_0(s;z_2,x,\bm{l}(s))-\Lambda_*(s;z_1,x)\} d\Lambda_0(s;z_2,x,\bm{l}(s))d\bm{l}(s) \\
&\quad + \int_0^t \int_{\mathcal{L}(s)} \prod_{0<r\leq s} q_0(l(r) \mid z_2,x,\bm{l}(r^-)) \exp\{-\Lambda_0(s;z_2,x,\bm{l}(s))-\Lambda_*(s;z_1,x)\} d\Lambda_*(s;z_1,x)d\bm{l}(s) \\
&\quad - \int_0^t \int_{\mathcal{L}(s)} \prod_{0<r\leq s} q_0(l(r) \mid z_2,x,\bm{l}(r^-)) \int_{\mathcal{L}(t)|\mathcal{L}(s)}\prod_{s<r\leq t} q_1(l(r) \mid z_2,s,x,\bm{l}(r^-)) \\
&\qquad \cdot \exp\{-\Lambda_0(s;z_2,x,\bm{l}(s))-\Lambda_*(s;z_1,x)-\Lambda_1(t;z_2,s,x,\bm{l}(t))\} d\Lambda_*(s;z_1,x)d\bm{l}(t),
\end{aligned} \label{est:haz_t}
\end{equation}
\end{small}
and hence $F(t;z_1,z_2) = \int_{\mathcal{X}}F(t;z_1,z_2,x)dP(x)$.
\end{theorem}

The proof of Theorem \ref{thm:id} is given in Supplementary Material A.3. The expression of $F(t;z_1,z_2,x)$ involves three terms. The first term is the conditional cumulative incidence of the first occurrence of the terminal event at time $t$ (without experiencing the intermediate event), integrating over the distribution of time-varying confounders. The second term is the conditional cumulative incidence of the intermediate event at time $t$, integrating over the distribution of time-varying confounders. The first occurrences of the terminal event and the intermediate event are a pair of competing events. The third term is the remaining probability of the intermediate event at time $t$, integrating over the distribution of the time-varying confounders. Note that the decline of the remaining probability of the intermediate event (i.e., the cumulative incidence of the indirect terminal event following the intermediate event) is a function of the density of the intermediate event, and there is an integral over time-varying confounders in the density of the intermediate event.

\subsection{Direct and indirect interventional effects}

We define the interventional direct effect (IDE) and interventional indirect effect (IIE) as
\[
\IDE(t;0) = F(t;0,1)-F(t;0,0), \ \IIE(t;1) = F(t;1,1)-F(t;0,1).
\]
If there are no time-varying confounders, then the counterfactual CIF $F(t;z,z)$ is identical to the potential CIF $F(t;z)$ under treatment assignment $z$, 
\begin{equation}
\begin{aligned}
F(t;z) &= 1 - \int_{\mathcal{X}}\exp\{-\Lambda_0(t;z,x)-\Lambda_*(t;z,x)\}dP(x) \\
&\quad - \int_{\mathcal{X}}\int_0^t \exp\{-\Lambda_0(s;z,x)-\Lambda_*(s;z,x)-\Lambda_1(t;z,s,x)\} d\Lambda_*(s;z,x)dP(x),
\end{aligned}
\end{equation}
which allows us to decompose the total effect into interventional direct and indirect effects. Details are provided in Supplementary Material A.1. However, in the presence of time-varying confounders, the counterfactual CIF $F(t;z,z)$ generally does not equal $F(t;z)$. The interventional direct and indirect effects sum to a so-called overall effect (OE),
\begin{align*}
\TE(t) &= F(t;1) - F(t;0) \\
\neq \OTE(t) &= F(t;1,1) - F(t;0,0) = \IDE(t;0) + \IIE(t;1).
\end{align*}

\begin{remark}
An alternative to Assumption \ref{ie:haz_t} is to assume that we draw the intermediate event process according to the prevalence,
\begin{equation}
\begin{aligned}
w_{n_1}(t;z_1,z_2,x,\bm{l}(t)) &:= P\{G(t;z_1)=n_1 \mid \tilde{N}_2(t^-;z_2,\bm{G}(t^-;z_1))=0, X=x, \bm{L}(t)=\bm{l}(t)\} \\
&= P\{\tilde{N}_1(t;z_1)=n_1 \mid \tilde{N}_2(t^-;z_1)=0, X=x\} =: w_{n_1}(t;z_1,x).
\end{aligned} \label{ie:prev_t}
\end{equation}
Like the hazard version draw, calculating the prevalence requires integrating $\bm{L}(t)$ out. However, it is not straightforward to model the prevalence. Usually, time-to-event data are modeled by hazards because the likelihood of observed data can be written as a function of hazards. The prevalence does not occur in the likelihood, so the hazard version draw is preferable over the prevalence version draw.
\end{remark}

\begin{remark} 
The randomized interventional approach does not satisfy the sharp meditational null criteria \cite{miles2023causal, diaz2024non}. Zheng and van der Laan \cite{zheng2017longitudinal} proposed defining the interventional effects conditional on the entire history of time-varying confounders. Instead of Assumption \ref{ie:haz_t}, the conditional version assumes
\begin{equation}
\begin{aligned}
d\Lambda_*(t;z_1,z_2,x,\bm{l}(t)) = d\Lambda_*(t;z_1,x,\bm{l}(t)),
\end{aligned} 
\end{equation}
so we do not need to integrate $\bm{L}(t)$ out. Following this assumption, the direct effect comprises the effect directly on the terminal event and the effect through the time-varying confounders on the terminal event. In contrast, the indirect effect is only through modifying the conditional hazard of the intermediate event on the terminal event. In this case, the conditional counterfactual cumulative incidence function is
\begin{small}
\begin{equation}
\begin{aligned}
F_{\CI}(t;z_1,z_2,x) &= \int_0^t \int_{\mathcal{L}(s)}\prod_{0<r\leq s} q_0(l(r) \mid z_2,x,\bm{l}(r^-)) \exp\{-\Lambda_0(s;z_2,x,\bm{l}(s))-\Lambda_*(s;z_1,x,\bm{l}(s))\} d\Lambda_0(s;z_2,x,\bm{l}(s))d\bm{l}(s) \\
&\quad + \int_0^t \int_{\mathcal{L}(s)} \prod_{0<r\leq s} q_0(l(r) \mid z_2,x,\bm{l}(r^-)) \exp\{-\Lambda_0(s;z_2,x,\bm{l}(s))-\Lambda_*(s;z_1,x,\bm{l}(s))\} d\Lambda_*(s;z_1,x,\bm{l}(s))d\bm{l}(s) \\
&\quad - \int_0^t \int_{\mathcal{L}(s)} \prod_{0<r\leq s} q_0(l(r) \mid z_2,x,\bm{l}(r^-)) \int_{\mathcal{L}(t)|\mathcal{L}(s)}\prod_{s<r\leq t} q_1(l(r) \mid z_2,s,x,\bm{l}(r^-)) \\
&\qquad \cdot \exp\{-\Lambda_0(s;z_2,x,\bm{l}(s))-\Lambda_*(s;z_1,x,\bm{l}(s))-\Lambda_1(t;z_2,s,x,\bm{l}(t))\} d\Lambda_*(s;z_1,x,\bm{l}(s))d\bm{l}(t).
\end{aligned}
\end{equation}
\end{small}
We observe that $F_{\CI}(t;z,z) = F(t;z)$, and the total effect is identical to the overall effect. Estimation is significantly easier under this assumption, as the counterfactual cumulative incidence is a simple combination of the conditional hazard terms in the likelihood. The conditional interventional direct direct effect (CDE) and conditional interventional indirect effect (CIE) are $\CDE(t;0) = F_{\CI}(t;0,1)-F(t;0), \ \CIE(t;1) = F(t;1)-F_{\CI}(t;0,1)$, respectively.
\end{remark}

\section{Estimation}

\subsection{Estimation based on the likelihood}

The likelihood function for a single individual $O = (X,Z,\bm{L}(T_2),T_1,\Delta_1,T_2,\Delta_2)$ is given by
\begin{align*}
f(O) &=  P(Z \mid X) \cdot \prod_{0<s\leq T_1} q_0(L(s)\mid Z,X,\bm{L}(s^-)) \prod_{T_1<s\leq T_2} q_1(L(s)\mid Z,T_1,X,\bm{L}(s^-)) \\
&\quad \cdot \exp\left\{-\Lambda_*(T_1;Z,X,\bm{L}(T_1))-\Lambda_0(T_1;Z,X,\bm{L}(T_1))\right\} \lambda_*(T_1;Z,X,\bm{L}(T_1))^{\Delta_1} \\
&\quad \cdot \exp\left\{-\Lambda_1(T_2;Z,T_1,X,\bm{L}(T_2))\right\} \lambda_{0}(T_2;Z,X,\bm{L}(T_2))^{(1-\Delta_1)\Delta_2} \lambda_{1}(T_2;Z,T_1,X,\bm{L}(T_2))^{\Delta_1\Delta_2},
\end{align*}
where $\lambda_*(\cdot)$, $\lambda_0(\cdot)$ and $\lambda_1(\cdot)$ are the derivatives of $\Lambda_*(\cdot)$, $\Lambda_0(\cdot)$ and $\Lambda_1(\cdot)$ with respect to the first argument, respectively. The product limits can be expressed as integration through logarithm--exponential transformation. If the parameters in $P(z\mid x)$, $Q_{n_1}(\cdot)$, $d\Lambda_{n_1}(\cdot)$, and $d\Lambda_*(\cdot)$ are variational independent, then the likelihood function can be factorized and maximized for each part separately. 

To illustrate the maximum likelihood estimation (MLE), we present a special case where the time-varying confounder $L(t) \in \{0,1\}$ is the counting process of an event. Assuming Markovness, we model the intensity of the time-varying confounder $L(t)$ by
\begin{align*}
q_0(l(t) \mid z,x,\bm{l}(t^-),\theta_{1z})dt &= [dQ(t;z)\exp\{(x,0)\beta_{1z}\}]^{I\{l(t^-)<l(t)\}} \cdot 1^{I\{l(t^-)=l(t)=1\}} \\
&\quad \cdot [1-dQ(t;z)\exp\{(x,0)\beta_{1z}\}]^{I\{l(t^-)=l(t)=0\}} \cdot 0^{I\{l(t^-)>l(t)\}}, \\
q_1(l(t) \mid z,r,x,\bm{l}(t^-),\theta_{1z})dt &= [dQ(t;z)\exp\{(x,1)\beta_{1z}\}]^{I\{l(t^-)<l(t)\}} \cdot 1^{I\{l(t^-)=l(t)=1\}} \\
&\quad \cdot [1-dQ(t;z)\exp\{(x,1)\beta_{1z}\}]^{I\{l(t^-)=l(t)=0\}} \cdot 0^{I\{l(t^-)>l(t)\}},
\end{align*}
where we define $0^0=1$. We model the hazard of the intermediate event by
\[
d\Lambda_*(t;z,x,\bm{l}(t),\theta_{2z}) = d\Lambda_*(t;z) \exp\{(x,l(t))\beta_{2z}\},
\]
and the cause-specific hazards of the terminal event by
\begin{align*}
d\Lambda_0(t;z,x,\bm{l}(t),\theta_{3z}) &= d\Lambda(t;z) \exp\{(x,l(t),0)\beta_{3z}\}, \\
d\Lambda_1(t;z,r,x,\bm{l}(t),\theta_{3z}) &= d\Lambda(t;z) \exp\{(x,l(t),1)\beta_{3z}\},
\end{align*}
where $\theta_{1z}=(Q(\cdot;z),\beta_{1z})$, $\theta_{2z}=(\Lambda_*(\cdot;z),\beta_{2z})$, and $\theta_{3z}=(\Lambda(\cdot;z),\beta_{3z})$ with $Q(\cdot;z)$, $\Lambda_*(\cdot;z)$ and $\Lambda(\cdot;z)$ being monotonically increasing functions with derivatives $q(\cdot;z)$, $\lambda_*(\cdot;z)$ and $\lambda(\cdot;z)$ on $[0,t^*]$ subject to $Q(0;z)=\Lambda_*(0;z)=\Lambda(0;z)=0$ and $Q(t^*;z)=\Lambda_*(t^*;z)=\Lambda(t^*;z)<M$ for a constant $M>0$. 

Given the sample $\mathcal{O} = \{O_i: i=1,\ldots,n\}$ with $n$ independent units, the infinitely-dimensional parameter $\theta=(\theta_{1z},\theta_{2z},\theta_{3z}: z=0,1)$ is estimated by the nonparametric maximum likelihood estimation (NPMLE), where the cumulative hazard (intensity) functions are estimated by step functions \cite{laird1978nonparametric, zeng2007maximum}. Specifically, let $\mathcal{S}$ be the set of step functions with jumps at event times $\{t: dL_i(t)=1\} \cup \{t: dN_{1i}(t)=1\} \cup \{t: dN_{2i}(t)=1\}$. Let $Q\{t\}$ be the jump size of a step function $Q(\cdot) \in \mathcal{S}$ at $t$. The model parameter (including nonparametric baseline hazards and parametric coefficients of confounders) is estimated by
\begin{align*}
\hat\theta_{1z} &= \argmax_{\beta\in\mathbb{R}^{p+1}, Q\in\mathcal{S}} \sum_{i:Z_i=z} \int_0^{T_{2i}}\left[\log(Q\{t\})+(X_i,N_{1i}(t))\beta\right] dL_i(t) -\int_0^{T_{2i}}I\{L_i(t)=0\}\exp\{(X_i,N_{1i}(t))\beta\}dQ(t), \\
\hat\theta_{2z} &= \argmax_{\beta\in\mathbb{R}^{p+1}, Q\in\mathcal{S}} \sum_{i:Z_i=z} \Delta_1\left[\log(Q\{T_{1i}\})+(X_i,L_{i}(T_{1i}))\beta\right] - \int_0^{T_{1i}} \exp\{(X_i,L_i(t))\beta\}dQ(t), \\ 
\hat\theta_{3z} &= \argmax_{\beta\in\mathbb{R}^{p+2}, Q\in\mathcal{S}} \sum_{i:Z_i=z} \Delta_2\left[\log(Q\{T_{2i}\})+(X_i,L_i(T_{2i}),N_{1i}(T_{2i}))\beta\right] - \int_0^{T_{2i}} \exp\{(X_i,L_i(t),N_{1i}(t))\beta\}dQ(t),
\end{align*}
where $p$ is the dimension of $X$. 

The estimator of $\theta$, denoted by $\hat\theta$, is consistent and converges to a Gaussian process with a root-$n$ convergence rate, as a standard result of the Cox model with time-varying covariates \cite{zeng2007maximum}. The conditional counterfactual CIF $F(t;z_1,z_2,x)$ is a function of $\theta$, so $F(t;z_1,z_2,x)$ can be estimated by plugging the maximum likelihood estimates into the formula, denoted by $\hat{F}(t;z_1,z_2,x)$. The population-level $F(t;z_1,z_2)$ is then estimated by the empirical average over the sample, $\hat{F}(t;z_1,z_2) = \mathbb{P}_n\{\hat{F}(t;z_1,z_2,X)\} := n^{-1}\sum_{i=1}^{n}\hat{F}(t;z_1,z_2,X_i)$. The resulting estimator for the counterfactual CIF $\hat{F}(t;z_1,z_2)$ is asymptotically normal and efficient. However, the expression of the asymptotic variance of $\hat{F}(t;z_1,z_2)$ is complicated. The confidence interval of $F(t;z_1,z_2)$ can be constructed by bootstrap. Technical details are provided in Supplementary Material D.

\subsection{Sensitivity analysis: incorporating a latent confounder}

Although the sequential ignorability in Assumption \ref{ie:haz2_t} allows time-varying confounders to moderate the hazards, it is doubtful whether there are unmeasured confounders in real-world studies. Generally, the counterfactual CIF is not nonparametrically identifiable in the presence of unmeasured confounders. In this section, we provide a trade-off between the nonparametric identifiability and unconfoundedness assumption by considering a latent frailty (random effect) $b$ that acts as an unmeasured confounder at baseline \cite{xu2010statistical}. The frailty $b$ captures the correlation between the hazards of the intermediate and terminal events, as well as the intensity of time-varying confounders. The frailty $b$ is assumed to be independent of baseline confounders $X$ and does not contribute to the treatment assignment or censoring mechanism, following a prespecified parametric distribution denoted by $\phi(b)$. We assume that a larger value of $b$ is associated with a higher risk of events, reflecting a vulnerable condition. 

Since $b$ is invariant over time, we can treat $b$ as a baseline variable, although unobserved. We assume latent sequential ignorability that the cause-specific hazards of the terminal event given $\bm{G}(\cdot)$ are identical to the natural hazards, which implicitly postulates that all confounding between $\tilde{N}_1(t)$ and $\tilde{N}_2(t)$ is adjusted by $X$, $\bm{L}(t)$, and $b$. The hazard of $G(\cdot;z_1)$ depends on baseline covariates including $X$ and $b$. An individual with a high value of $b$ is inherently vulnerable to the intermediate event; therefore, the hazard according to which to draw the intermediate event should also be large, reflecting its vulnerable nature.
The counterfactual CIF $F(t;z_1,z_2) = \int_{b}\int_{\mathcal{X}} F(t;z_1,z_2,x,b) \phi(b)dP(x)db$.

When incorporating frailty, the identifiability of functions in the likelihood relies on the specific model chosen and requires a case-by-case justification. We assume that the frailty $b \sim N(0,1)$ with density $\phi(b)=(2\pi)^{-1/2}\exp(-b^2/2)$. We still impose proportional hazards models for $Q(\cdot)$, $\Lambda_*(\cdot)$, and $\Lambda(\cdot)$ by assuming that the hazards ratio of $b$ is $\exp(\alpha_z)$ in the treatment group $z$. We assume that $\alpha_z>0$ and the components in $X$ and $b$ are not completely linearly correlated for identifiability.
The likelihood function for a single individual is given by (omitting the propensity score and censoring distribution since they are nuisance)
\begin{align*}
f(O) &= \int_{-\infty}^{+\infty} \exp\left\{-\int_0^{T_2}I\{L_i(t)=0\}e^{(X,N_1(t))\beta_{1Z}+\alpha_Zb}dQ(t;Z)\right\} \prod_{0<s\leq T_2}\left[ q(s;Z)e^{(X,N_i(s))\beta_{1Z}+\alpha_Zb}\right]^{dL(s)} \\
&\quad \cdot \exp\left\{-\int_0^{T_1}e^{(X,L(t))\beta_{2Z}+\alpha_Zb}d\Lambda_*(t;Z)\right\} \left[\lambda_*(T_1;Z)e^{(X,L(T_1))\beta_{2Z}+\alpha_Zb}\right]^{\Delta_1} \\
&\quad \cdot \exp\left\{-\int_0^{T_2}e^{(X,L(t),N_1(t))\beta_{3Z}+\alpha_Zb}d\Lambda(t;Z)\right\} \left[\lambda(T_3;Z)e^{(X,L(T_2),N_1(T_2))\beta_{3Z}+\alpha_Zb}\right]^{\Delta_2} \phi(b)db.
\end{align*}

Directly estimating the parameters by optimization is computationally inefficient due to integration over the frailty. Provided that the parameter $(\theta,\alpha_z: z=0,1)$ is identifiable, the Expectation--Maximization (EM) algorithm can be employed to estimate the parameters by regarding $b$ as missing data. In the E-step, we calculate the conditional expectation of any function $g(O,b)$ given observed data at current parameter values. In the M-step, we update parameters by the conditional expectation of the complete-data log-likelihood. Specifically, we update the parametric part using a one-step Newton--Raphson method for the partial-likelihood score function, and then update the nonparametric part using Breslow estimators. To further improve computational efficiency, an alternative to the EM algorithm is to use the Laplace approximation to approximate the integration \cite{therneau2003penalized, ripatti2004estimation}. Finally, the target causal estimand $F(t;z_1,z_2)$ can be estimated by the empirical average of the conditional estimate for each individual by plugging the obtained likelihood-based estimators into the identification formula and integrating $b$ out, $\hat{F}(t;z_1,z_2) = \int_{-\infty}^{+\infty}\mathbb{P}_n[\hat{F}(t;z_1,z_2,X,b)]\phi(b)db$. The detailed estimation procedure is given in Supplementary Material E. Under some regularity conditions to ensure identifiability, the resulting estimators $(\hat\theta,\hat\alpha_0,\hat\alpha_1)$ and $\hat{F}(t;z_1,z_2)$ are consistent, asymptotically normal, and asymptotically efficient \cite{zeng2009semiparametric}. Therefore, bootstrap can be applied for inference.

Since the baseline hazards are nonparametric, the frailty effects are weakly identified. The variation of frailty may be absorbed into the baseline hazards, resulting in underestimated frailty effects in a finite sample. Some studies also indicate that the estimation is insensitive to the misspecification of the frailty distribution \cite{gasparini2019impact}. Nevertheless, frailty modeling can yield slightly more accurate cause-specific hazard estimates than marginal models in the presence of unmeasured confounders, as it allows an additional degree of freedom \cite{faraggi1996competing, troendle2018dealing}. Some special cases of time-varying confounders favor alternative estimation methods. Suppose the time-varying confounders are longitudinal, with measurements at fixed points. In that case, we can use the $g$-formula to express the counterfactual CIF by adding a confounder transition intensity at the monitoring times. Suppose the time-varying confounders for each subject are possibly measured at different times. In that case, we may use the last observation carried forward (LOCF) to impute the time-varying confounders throughout the study period.

\section{Simulation studies}

In this section, we conduct simulation studies to assess the performance of likelihood-based estimators of counterfactual cumulative incidence and randomized interventional effects. 
To mimic the sample size in the data application, we generate a sample of size $n=600$. We generate three independent baseline covariates $X=(X_1,X_2,X_3)$, each following a Bernoulli distribution $B(0.5)$. To mimic the GVHD in the real-data application, we assume there is a time-varying confounder $L(t)$ that can jump from 0 to 1 at some time. This time-varying confounder can modify the hazards of the intermediate event (relapse) and the terminal event (death) in opposite directions. The hazard of this time-varying confounder at time $t$, when the value of the intermediate event counting process is $n_1$, is assumed to be
\[
\lambda^*_{n_1}(t;z,x) = (0.01+0.04t-0.02zt)\exp(0.2x_1+0.2x_2-0.2x_3-0.2n_1+U). 
\]
Thus, $q_0(l(t)=1\mid z,x,l(t^-)=0) = \lambda_0^*(t;z,x)$ and $q_1(l(t)=1\mid z,r,x,l(t^-)=0) = \lambda_1^*(t;z,x)$. Under Markovness, let the hazards of the intermediate event and terminal event be
\begin{align*}
\lambda_*(t;z,x,\bm{l}(t)) &= (0.02+0.02t+0.01zt) \exp(0.2x_1-0.2x_2-0.2x_3-0.3l(t)+U), \\
\lambda_{n_1}(t;z,x,\bm{l}(t)) &= (0.01+0.01z+0.02t) \exp(0.1x_1+0.1x_2+0.1x_3+0.5n_1+0.1(1+z)l(t)+U).
\end{align*}
We generate the censoring time from the hazard
\begin{align*}
\lambda_c(t;z=1,x,\bm{l}(t)) &= \{0.05(t-5)\}^2\exp(-0.1x_1), \\
\lambda_c(t;z=0,x,\bm{l}(t)) &= \{0.04(t-6)\}^2\exp(-0.1x_1-0.2x_2),
\end{align*}
and set the maximum follow-up time to be 15. The probability of receiving treatment is
\[
P(Z=1 \mid X=x) = 1/\{1+\exp(0.5+0.3x_1-0.4x_2-0.5x_3)\}.
\]

We consider three settings. The first setting is $U=0$ so that unconfoundedness and sequential ignorability hold. The median follow-up time is 6.8. About 68\% of individuals experience the terminal event, and 41\% experience the intermediate event before loss of follow-up. The second setting is $U \sim N(0,1)$ so that the frailty effect is 1 in each group. The third setting is $U \sim \log\{\mathrm{Gamma}(4,4)\}$. The median follow-up time and proportion of individuals who experience events in the last two settings are similar to those in the first setting. We generate $B=1000$ independent samples to estimate the cumulative incidences and treatment effects. Standard errors are obtained by a bootstrap with 200 resamplings. The true values are calculated by numerical integration using the identification formula.

For estimation, we consider NPMLE without and with frailty. Specifically, the model parameter is estimated using the `survival' package for marginal models and the `coxme' package (which uses the Laplace approximation) for frailty models in R. In our data-generating process, all events can be written as counting processes, so they can be reformulated as a multi-state model. The cumulative incidence of the terminal event is contributed by five paths: (1) initial state to terminal event, (2) initial state to time-varying confounder to terminal event, (3) initial state to intermediate event to terminal event, (4) initial state to time-varying confounder to intermediate event to terminal event, and (5) initial state to intermediate event to time-varying confounder to terminal event. Therefore, we can use the Kolmogorov forward equation to derive the conditional cumulative incidence function as a function of event-specific hazards \cite{andersen2002multi}.

The simulation is conducted using R 4.4.3 on a Linux server with an Intel(R) Xeon(R) Gold 6140, 2.30 GHz CPU. The median computation time to obtain the point estimates is 4 seconds for the model without frailty and 31 seconds for the model with frailty. In Supplementary Material G, we plot the true and estimated counterfactual CIFs. The direct effects (IDE and CDE) grow larger than 0.10 by $t=10$, while the true indirect effects (IIE and CIE) are always near zero. Table \ref{tab:bias2} shows the bias, standard deviation (SD), average standard error (SE) of the point estimates, and coverage percentage (CP) of nominal 95\% bootstrap confidence intervals at selected time points in these $B=1000$ samples. For illustration, we multiply the bias, SD, and SE by 100. In Setting 1, unconfoundedness holds, and both estimates have negligible bias. The relative bias of the estimated IDE and CDE is smaller than 5\%, and the absolute bias is smaller than 0.01. In Setting 2 with a normal frailty, the frailty model yields a slightly lower bias. In Setting 3 with a gamma frailty, both models are misspecified, and we observe that the frailty model yields a slightly lower bias. The confidence intervals associated with both methods can achieve the nominal coverage rate. In fact, the nonparametric baseline hazards can absorb part of the frailty effect, so the bias of the model without frailty is small even if the frailty is not modeled or the frailty distribution is misspecified. This phenomenon is consistent with observations made in other works \cite{troendle2018dealing, gasparini2019impact}. Due to the flexibility of the semiparametric model, the frailty effect is only weakly identified, yielding unstable estimates. Although the frailty model yields point estimates with slightly smaller bias, the bootstrap standard error can be large. As a result, the confidence intervals from the frailty model show slight over-coverage.

\begin{table}
\centering
\caption{Bias, standard deviation (SD), average standard error (SE) of the estimated treatment effects multiplied by 100, and coverage percentage (CP) of nominal 95\% confidence intervals at selected time points in the $B=1000$ samples} \label{tab:bias2}
\begin{tabular}{llrrrrrrrrrr}
  \toprule
  & & \multicolumn{5}{c}{Model without frailty} & \multicolumn{5}{c}{Model with frailty} \\
  \cmidrule(lr){3-7} \cmidrule(lr){8-12}
  & $t$ & 2 & 4 & 6 & 8 & 10 & 2 & 4 & 6 & 8 & 10 \\ 
  \midrule
  \multicolumn{12}{l}{Setting 1: $U= 0$} \\
  IDE & Bias & -0.059 & -0.149 & -0.492 & -0.310 & -0.622 & -0.053 & -0.126 & -0.434 & -0.211 & -0.515 \\ 
  & SD & 2.248 & 3.330 & 3.907 & 4.046 & 4.255 & 2.248 & 3.329 & 3.902 & 4.040 & 4.248 \\ 
  & SE & 2.223 & 3.391 & 3.959 & 4.097 & 4.066 & 2.214 & 3.381 & 3.964 & 4.275 & 4.980 \\ 
  & CP & 0.957 & 0.951 & 0.948 & 0.950 & 0.932 & 0.956 & 0.951 & 0.949 & 0.953 & 0.941 \\ 
  IIE & Bias & -0.001 & -0.007 & -0.028 & -0.063 & -0.119 & -0.005 & -0.028 & -0.086 & -0.165 & -0.235 \\ 
  & SD & 0.085 & 0.300 & 0.542 & 0.666 & 0.618 & 0.081 & 0.287 & 0.511 & 0.615 & 0.568 \\ 
  & SE & 0.092 & 0.329 & 0.565 & 0.657 & 0.582 & 0.091 & 0.328 & 0.625 & 1.004 & 1.709 \\ 
  & CP & 0.982 & 0.971 & 0.953 & 0.940 & 0.910 & 0.983 & 0.972 & 0.952 & 0.929 & 0.906 \\ 
  CDE & Bias & -0.059 & -0.150 & -0.493 & -0.311 & -0.626 & -0.053 & -0.123 & -0.426 & -0.200 & -0.512 \\ 
  & SD & 2.248 & 3.330 & 3.907 & 4.043 & 4.251 & 2.248 & 3.329 & 3.901 & 4.037 & 4.241 \\ 
  & SE & 2.223 & 3.391 & 3.959 & 4.095 & 4.062 & 2.214 & 3.382 & 3.964 & 4.273 & 4.949 \\ 
  & CP & 0.957 & 0.951 & 0.948 & 0.951 & 0.932 & 0.956 & 0.951 & 0.949 & 0.953 & 0.941 \\ 
  CIE & Bias & -0.001 & -0.007 & -0.027 & -0.061 & -0.116 & -0.005 & -0.031 & -0.096 & -0.180 & -0.248 \\ 
  & SD & 0.085 & 0.299 & 0.538 & 0.660 & 0.613 & 0.081 & 0.286 & 0.506 & 0.610 & 0.561 \\ 
  & SE & 0.092 & 0.328 & 0.563 & 0.654 & 0.578 & 0.091 & 0.321 & 0.549 & 0.679 & 0.772 \\ 
  & CP & 0.983 & 0.974 & 0.960 & 0.949 & 0.929 & 0.985 & 0.978 & 0.964 & 0.945 & 0.926 \\ 
  \midrule
  \multicolumn{12}{l}{Setting 2: $U \sim N(0,1)$} \\
  IDE & Bias & -0.244 & -0.408 & -0.446 & -0.420 & -0.613 & -0.244 & -0.373 & -0.328 & -0.232 & -0.620 \\ 
  & SD & 2.830 & 3.693 & 3.979 & 3.939 & 4.235 & 2.838 & 3.722 & 3.989 & 4.719 & 5.833 \\ 
  & SE & 2.646 & 3.602 & 3.896 & 3.946 & 4.082 & 2.633 & 3.735 & 4.787 & 6.386 & 8.511 \\ 
  & CP & 0.924 & 0.944 & 0.952 & 0.953 & 0.938 & 0.922 & 0.949 & 0.962 & 0.971 & 0.974 \\ 
  IIE & Bias & 0.103 & 0.334 & 0.376 & 0.260 & 0.080 & 0.111 & 0.302 & 0.247 & 0.038 & -0.047 \\ 
  & SD & 0.329 & 0.826 & 1.120 & 1.211 & 1.186 & 0.320 & 0.774 & 1.030 & 1.106 & 3.425 \\ 
  & SE & 0.348 & 0.835 & 1.114 & 1.186 & 1.148 & 0.349 & 0.970 & 1.882 & 3.198 & 4.999 \\ 
  & CP & 0.981 & 0.957 & 0.952 & 0.959 & 0.961 & 0.982 & 0.969 & 0.972 & 0.981 & 0.982 \\ 
  CDE & Bias & -0.268 & -0.532 & -0.703 & -0.776 & -1.008 & -0.263 & -0.458 & -0.493 & -0.505 & -0.772 \\ 
  & SD & 2.831 & 3.693 & 3.976 & 3.931 & 4.226 & 2.839 & 3.725 & 3.993 & 4.409 & 6.139 \\ 
  & SE & 2.647 & 3.602 & 3.895 & 3.947 & 4.086 & 2.635 & 3.742 & 4.853 & 6.644 & 9.203 \\ 
  & CP & 0.925 & 0.942 & 0.953 & 0.952 & 0.937 & 0.922 & 0.945 & 0.969 & 0.975 & 0.978 \\ 
  CIE & Bias & 0.125 & 0.436 & 0.572 & 0.527 & 0.380 & 0.129 & 0.373 & 0.373 & 0.200 & -0.021 \\ 
  & SD & 0.331 & 0.827 & 1.116 & 1.212 & 1.205 & 0.322 & 0.781 & 1.052 & 1.146 & 1.143 \\ 
  & SE & 0.350 & 0.836 & 1.111 & 1.185 & 1.158 & 0.341 & 0.843 & 1.305 & 1.899 & 2.782 \\ 
  & CP & 0.978 & 0.949 & 0.938 & 0.943 & 0.947 & 0.982 & 0.963 & 0.967 & 0.973 & 0.980 \\ 
  \midrule
  \multicolumn{12}{l}{Setting 3: $U \sim \log\{\mathrm{Gamma}(4,4)\}$} \\
  IDE & Bias & -0.125 & -0.341 & -0.571 & -0.321 & -0.529 & -0.117 & -0.314 & -0.503 & -0.205 & -0.394 \\ 
  & SD & 2.226 & 3.337 & 3.940 & 4.242 & 4.505 & 2.226 & 3.339 & 3.946 & 4.243 & 4.496 \\ 
  & SE & 2.208 & 3.341 & 3.904 & 4.101 & 4.271 & 2.198 & 3.329 & 3.912 & 4.271 & 4.985 \\ 
  & CP & 0.938 & 0.950 & 0.941 & 0.939 & 0.936 & 0.937 & 0.950 & 0.939 & 0.941 & 0.942 \\ 
  IIE & Bias & 0.022 & 0.066 & 0.104 & 0.096 & 0.025 & 0.020 & 0.050 & 0.047 & -0.013 & -0.113 \\ 
  & SD & 0.121 & 0.408 & 0.708 & 0.847 & 0.832 & 0.117 & 0.391 & 0.670 & 0.788 & 0.771 \\ 
  & SE & 0.133 & 0.441 & 0.732 & 0.861 & 0.836 & 0.128 & 0.423 & 0.780 & 1.132 & 1.696 \\ 
  & CP & 0.987 & 0.986 & 0.977 & 0.967 & 0.971 & 0.985 & 0.986 & 0.977 & 0.971 & 0.974 \\ 
  CDE & Bias & -0.126 & -0.356 & -0.613 & -0.385 & -0.598 & -0.117 & -0.321 & -0.520 & -0.225 & -0.409 \\ 
  & SD & 2.226 & 3.338 & 3.942 & 4.244 & 4.506 & 2.226 & 3.341 & 3.951 & 4.249 & 4.496 \\ 
  & SE & 2.208 & 3.341 & 3.904 & 4.102 & 4.272 & 2.198 & 3.330 & 3.913 & 4.271 & 4.975 \\ 
  & CP & 0.938 & 0.950 & 0.939 & 0.941 & 0.933 & 0.937 & 0.950 & 0.940 & 0.943 & 0.942 \\ 
  CIE & Bias & 0.024 & 0.081 & 0.146 & 0.157 & 0.088 & 0.021 & 0.058 & 0.065 & 0.006 & -0.101 \\ 
  & SD & 0.121 & 0.407 & 0.710 & 0.856 & 0.847 & 0.117 & 0.393 & 0.679 & 0.811 & 0.800 \\ 
  & SE & 0.133 & 0.441 & 0.733 & 0.865 & 0.844 & 0.128 & 0.422 & 0.706 & 0.852 & 1.040 \\ 
  & CP & 0.989 & 0.985 & 0.978 & 0.970 & 0.972 & 0.987 & 0.987 & 0.982 & 0.976 & 0.970 \\ 
  \bottomrule
\end{tabular}
\end{table}

\section{Application to a hematopoietic cell transplantation study}

\subsection{Data description}

Hematopoietic cell transplantation (HCT) is a well-recognized approach to treating lymphoma. Due to mismatches in human leukocyte antigen (HLA) loci between the donor and recipient, graft-versus-host disease (GVHD) often occurs after transplantation. GVHD is a significant source of transplant-related mortality. HLA-matched related donor transplantation has long been regarded as the first choice of transplantation because there will be the least GVHD after transplantation. Other common transplant modalities include matched unrelated donor (MUD) transplantation from unrelated donors and haploidentical (Haplo) hematopoietic cell transplantation from related donors. Previous studies found that MUD-HCT is associated with higher risks of GVHD and death compared to Haplo-HCT. In contrast, the difference in relapse risk is insignificant \cite{grunwald2021alternative}. Therefore, in the absence of HLA-matched related donor transplantation, international guidelines suggested using haploidentical hematopoietic cell transplantation due to its lower GVHD risk compared with unrelated donor transplantation. 

In the past few decades, the use of post-transplant cyclophosphamide (PTCy)-based GVHD prophylaxis has rapidly expanded, as it can reduce the risk of GVHD for haploidentical transplantation \cite{luznik2008hla}. Some studies found that PTCy-based GVHD prophylaxis is also beneficial for matched unrelated donor transplantation \cite{gooptu2021hla, mussetti2023haploidentical}. Since the risk of GVHD undergoing MUD-HCT can be lowered by prophylaxis, the transplant-related mortality undergoing MUD-HCT is reduced. After properly controlling the risk of GVHD, MUD-HCT has the potential to achieve a comparable prognosis compared to Haplo-HCT. Therefore, it is desired to compare the performance of MUD-HCT and Haplo-HCT when using PTCy-based GVHD prophylaxis. The scientific questions lie in two aspects. Does MUD-HCT lead to lower transplant-related mortality (direct effect)? Does MUD-HCT lead to lower relapse-related mortality (indirect effect)? Since GVHD has opposite effects on relapse and death simultaneously, we need to envision a hypothetical world in which the GVHD effects on these two events can be controlled separately. This motivates us to consider hypothetical interventions in relapse, the intermediate event, so that its hazard can be controlled in a hypothetical world.

The cohort includes 760 adult patients with Hodgkin lymphoma (HL) or non-Hodgkin lymphoma (NHL) treated with Haplo-HCT ($Z=0$) or MUD-HCT ($Z=1$) using PTCy-based GVHD prophylaxis between 2010 and 2019 in the Center for International Blood and Marrow Transplant Research (CIBMTR) \cite{mussetti2023haploidentical}. Recipients of Haplo-HCT were mismatched at two or more HLA loci, whereas MUD transplants were matched at the allele level. We control six covariates that were collected in the cohort: age, sex (male or female), Karnofsky performance score ($\ge$90 or $<$90), type of GVHD prophylaxis (PTCy+CNI+MMF or PTCy+others), disease type (HL or NHL), and graft type (marrow or PBSC). We exclude individuals with missing Karnofsky performance score data, resulting in a sample of 636 individuals. Among these patients, 528 (83.0\%) received Haplo-HCT and 108 (17.0\%) received MUD-HCT. Table \ref{tab:sum} presents the summary statistics of baseline covariates in the sample.

\begin{table}
\centering
\caption{Summary statistics of the sample, stratified by treatment groups (continuous variables are tested by two-sided $t$-test and binary variables are tested by Chi-squared test)} \label{tab:sum}
\begin{tabular}{lrrrrr}
  \toprule
 & \multicolumn{2}{c}{Haplo-HCT ($Z=0$)} & \multicolumn{2}{c}{MUD-HCT ($Z=1$)} & \multirow{2}{*}{$P$-value} \\ 
 \cmidrule(lr){2-3} \cmidrule(lr){4-5}
 & Mean/Count & (SD/Prop) & Mean/Count & (SD/Prop) \\
  \midrule
  Age & 47.9 & (14.7) & 51.6 & (16.0) & 0.014 \\ 
  \multicolumn{6}{l}{Sex} \\
  ~~~~`Male' = 1 & 341 & (64.6\%) & 82 & (75.9\%) & 0.030 \\ 
  \multicolumn{6}{l}{Karnofsky performance score (Karnofcat)} \\
   ~~~~`$\ge90$' = 1 & 332 & (62.9\%) & 73 & (67.6\%) & 0.413 \\ 
  \multicolumn{6}{l}{Type of GVHD prophylaxis (Gvhprhrx)} \\
   ~~~~`PTCy+CNI+MMF' = 1 & 589 & (92.6\%) & 67 & (62.0\%) & $<$0.001 \\ 
  \multicolumn{6}{l}{Disease type (Disease)} \\
  ~~~~`NHL' = 1 & 347 & (65.7\%) & 80 & (74.1\%) & 0.116 \\ 
  \multicolumn{6}{l}{Graft type (Graftype)} \\
  ~~~~`Marrow' = 1 & 233 & (44.1\%) & 18 & (16.7\%) & $<$0.001 \\
   \bottomrule
\end{tabular}
\end{table}

The maximum follow-up time is 123.0 months, and the median follow-up time is 23.8 months. The intermediate event is relapse, and the terminal event is death. The time-varying confounder is the status of GVHD, including acute GVHD of levels II to IV and chronic GVHD. GVHD can occur either before or after relapse. Standard Cox regression indicates that MUD-HCT has a lower hazard of GVHD ($P<0.001$) and a lower hazard of all-cause death ($P=0.044$). The difference in the hazards of relapse is insignificant between groups ($P=0.994$).

\subsection{Modeling with GVHD as a time-varying confounder}

To answer the causal questions, we evaluate three counterfactual CIFs of death: $F(t;0,0)$, $F(t;1,1)$, and $F(t;0,1)$. We consider two types of estimands in this study. The first is the interventional effects estimand, in which we assume that the marginal hazard of relapse in survivors in the hypothetical world is identical to that in the Haplo-HCT group (marginalizing over GVHD status). In contrast, the hazards of other events remain natural. The second is the conditional interventional effects estimand, in which we hold the conditional hazard of relapse in the hypothetical world at the level observed under Haplo-HCT (conditional on GVHD status). This conditional version of the estimand can also be interpreted as natural effects (path-specific effects) by making an intervention on one of the two sets of paths: all paths leading to relapse and all other paths. The draw of the relapse process is unconditional on GVHD for the first estimand and conditional on GVHD for the second estimand.

We impose semiparametric Cox proportional hazards models for the transition rates of GVHD $d\Lambda^*_{n_1}(t;z,x)$, relapse $d\Lambda_{*}(t;z,x,l)$, and death $d\Lambda_{n_1}(t;z,x,l)$ at time $t$ assuming Markovness,
\begin{align*}
d\Lambda^*_{n_1}(t;z,x) &= d\Lambda_{0,z}(t) \exp(\beta_{0,z}' x + \eta_{0,z}n_1), \\
d\Lambda_{*}(t;z,x,l) &= d\Lambda_{1,z}(t) \exp(\beta_{1,z}' x + \gamma_{1,z} l), \\
d\Lambda_{n_1}(t;z,x,l) &= d\Lambda_{2,z}(t) \exp(\beta_{2,z}' x + \gamma_{2,z} l + \eta_{2,z}n_1),
\end{align*}
where $\Lambda_{0,z}(t)$, $\Lambda_{1,z}(t)$, and $\Lambda_{2,z}(t)$ are unknown baseline cumulative hazard functions, which can be different across treatment groups. The statuses of relapse and GVHD serve as time-varying covariates in the Cox model. The parameters in the above models are estimated using nonparametric maximum likelihood estimation (NPMLE), where the baseline hazards are modeled as step functions, as described in the previous section. Details of the estimation procedure and asymptotic properties are provided in Supplementary Material D. Estimates of coefficients and standard errors in the Cox models are displayed in Panel (A) of Table \ref{tab:est1}. The coefficients are the logarithms of the hazard ratios (HRs). The proportional hazards assumption is tested using Schoenfeld residuals \cite{Grambsch1994proportional}. The proportional hazards assumption for the event indicators as time-varying covariates is not rejected. 
GVHD reduces the risk of relapse, particularly with Haplo-HCT ($Z=0$), a phenomenon known as the ``graft-versus-leukemia'' effect \cite{kolb2008graft}. GVHD increases the risk of death; however, due to the limited sample size of the MUD-HCT ($Z=1$) group, the coefficient is not significant. Relapse greatly increases the risk of death.

\begin{table}
\centering
\caption{Estimated coefficients and standard errors in the Cox models for GVHD, relapse, and death in the data application} \label{tab:est1}
\begin{tabular}{lcccccccc}
  \toprule
  Event & \multicolumn{2}{c}{GVHD} & \multicolumn{2}{c}{Relapse} & \multicolumn{2}{c}{Death}  \\
  \cmidrule(lr){1-1} \cmidrule(lr){2-3} \cmidrule(lr){4-5} \cmidrule(lr){6-7}
  Group & MUD-HCT & Haplo-HCT & MUD-HCT & Haplo-HCT & MUD-HCT & Haplo-HCT \\
  \midrule
  \multicolumn{7}{l}{(A) Model without frailty} \\
  Age & -0.001 (0.016) & 0.005 (0.005) & -0.069 (0.019) & -0.009 (0.007) & 0.040 (0.023) & 0.016 (0.007) \\ 
  Sex & -0.186 (0.346) & -0.193 (0.122) & 0.427 (0.456) & -0.094 (0.171) & -1.312 (0.563) & 0.046 (0.172) \\ 
  Karnofcat & -0.548 (0.397) & 0.109 (0.125) & 0.497 (0.385) & 0.261 (0.185) & -0.647 (0.435) & 0.264 (0.161) \\ 
  Gvhprhrx & 0.712 (0.319) & -0.246 (0.218) & 0.177 (0.528) & -0.008 (0.343) & 0.759 (0.471) & -0.261 (0.334) \\ 
  Disease & -0.056 (0.513) & -0.233 (0.165) & 2.079 (0.622) & 0.357 (0.228) & -1.031 (0.781) & 0.027 (0.202) \\ 
  Graftype & 0.450 (0.407) & -0.423 (0.122) & 0.728 (0.512) & 0.410 (0.176) & -0.558 (0.643) & -0.063 (0.161) \\ 
  GVHD ($\hat\gamma$) & --- & --- & -0.172 (0.478) & -0.340 (0.175) & 0.696 (0.414) & 0.677 (0.156) \\ 
  Relapse ($\hat\eta$) & -0.149 (0.812) & 0.156 (0.271) & --- & --- & 3.414 (0.485) & 1.930 (0.167) \\ 
  \midrule
  \multicolumn{7}{l}{(B) Model with frailty} \\
  Age & -0.001 (0.015) & 0.005 (0.006) & -0.069 (0.019) & -0.009 (0.008) & 0.040 (0.021) & 0.017 (0.007) \\ 
  Sex & -0.186 (0.366) & -0.223 (0.136) & 0.427 (0.506) & -0.131 (0.183) & -1.312 (0.652) & 0.064 (0.172) \\ 
  Karnofcat & -0.548 (0.363) & 0.152 (0.136) & 0.497 (0.398) & 0.284 (0.186) & -0.647 (0.499) & 0.299 (0.170) \\ 
  Gvhprhrx & 0.712 (0.332) & -0.295 (0.256) & 0.177 (0.485) & -0.043 (0.349) & 0.759 (0.499) & -0.288 (0.338) \\ 
  Disease & -0.056 (0.511) & -0.237 (0.178) & 2.079 (0.728) & 0.381 (0.240) & -1.031 (0.764) & 0.052 (0.232) \\ 
  Graftype & 0.450 (0.411) & -0.449 (0.133) & 0.728 (0.514) & 0.369 (0.180) & -0.558 (0.676) & -0.116 (0.169) \\ 
  GVHD ($\hat\gamma$) & --- & --- & -0.173 (0.438) & -0.715 (0.178) & 0.695 (0.454) & 0.346 (0.171) \\
  Relapse ($\hat\eta$) & -0.149 (0.766) & -0.035 (0.266) & --- & --- & 3.414 (0.565) & 1.795 (0.169) \\ 
  \bottomrule
\end{tabular}
\end{table}

Since the time-varying confounder is a counting process, the data can be reformulated as a multi-state model with four states: initial state, GVHD state, relapse state, and death state. GVHD and relapse can transition to each other, but each state can only be visited once. The estimands under comparison can all be expressed as functions of the three hazards above; see Supplementary Material C. We plug the fitted models to obtain empirical estimates of the counterfactual CIF averaged over the sample. The computation takes 6 seconds on a laptop with a 13th Gen Intel(R) Core(TM) i7-1365U, 1800 MHz CPU. We then use a bootstrap with 200 resamplings to obtain the standard error. The 95\% bootstrap confidence interval is constructed based on the point estimates and standard errors.

\begin{figure}
\centering
\includegraphics[width=0.95\textwidth]{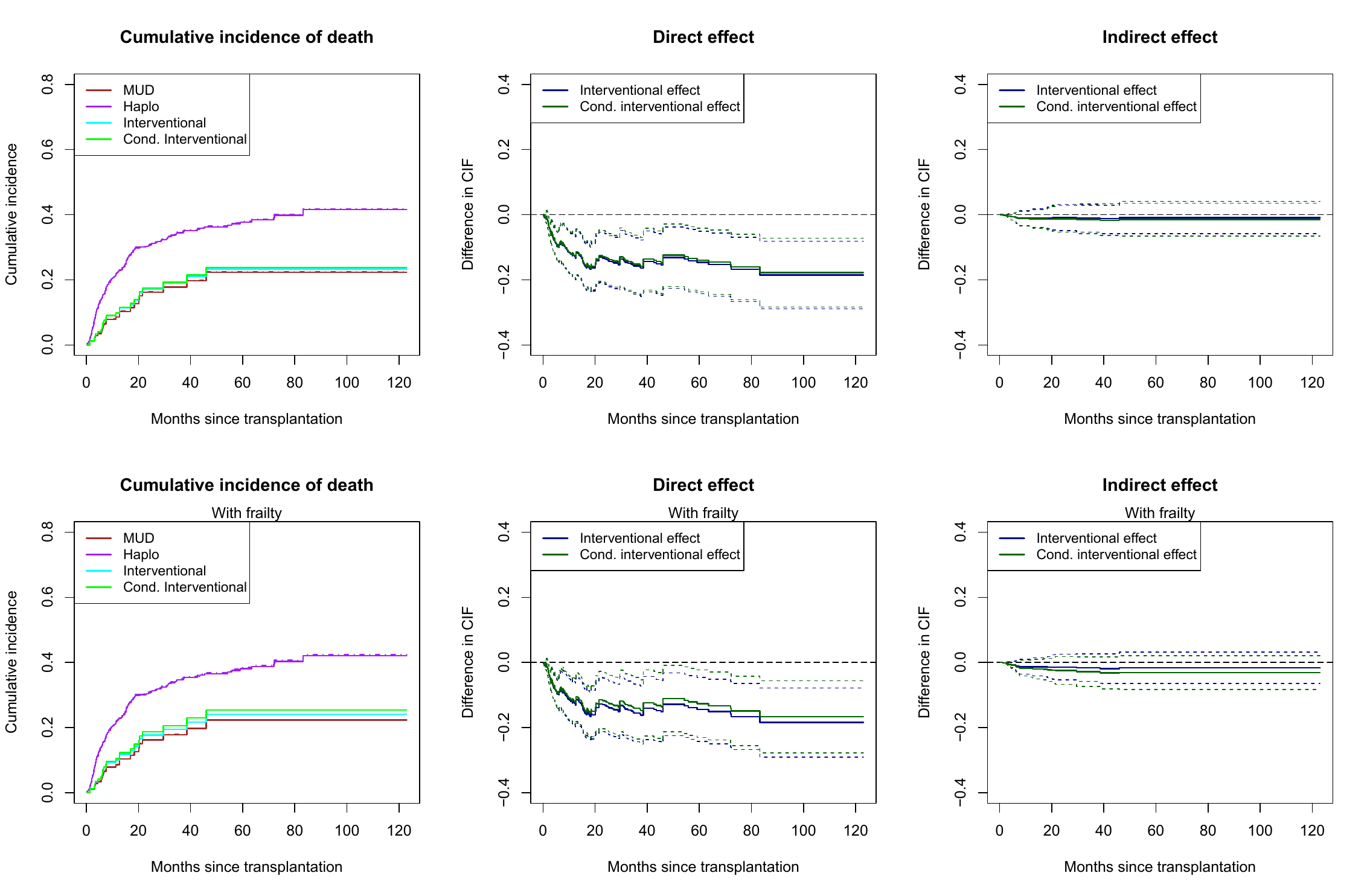}
\caption{Estimated cumulative incidence functions of death associated with Haplo-HCT ($z=0$), MUD-HCT ($z=1$), and in the hypothetical worlds ($z_1=0$, $z_2=1$); estimated interventional effects and conditional interventional effects. In the first column, the solid brown/purple lines are the estimated potential CIFs $\hat{F}(t;z)$, and the dashed lines are the estimated counterfactual CIFs $\hat{F}(t;z,z)$: they are very close to each other. Confidence intervals are obtained by bootstrapping. Upper: model without frailty; Lower: model with frailty.} \label{fig:cif1}
\end{figure}

The upper row of Figure \ref{fig:cif1} shows the estimated CIFs of death associated with Haplo-HCT $\hat{F}(t;0)$, MUD-HCT $\hat{F}(t;1)$, and in the hypothetical world $\hat{F}(t;0,1)$ in solid lines. For the interventional effects approach, note that the counterfactual CIF $F(t;z,z)$ is not identical to the potential CIF $F(t;z)$. We display the estimated counterfactual CIFs $\hat{F}(t;0,0)$ and $\hat{F}(t;1,1)$ in dashed lines. The estimated counterfactual CIFs are very close to the estimated potential CIFs, although not identical. We notice that the hazard of GVHD is much lower in the MUD-HCT group than in the Haplo-HCT group. Since GVHD is the main cause of death, MUD-HCT leads to lower mortality. Figure \ref{fig:cif1} also shows the estimated direct and indirect effects. The direct effect is significantly negative, whereas the indirect effect is close to zero.

\subsection{Modeling with a time-varying confounder and frailty}

We impose the following proportional hazards models incorporating the frailty $b$,
\begin{align*}
d\Lambda^*_{n_1}(t;z,x,b) &= d\Lambda_{0,z}(t) \exp(\beta_{0,z}' x + \eta_{0,z}n_1 + \alpha_z b), \\
d\Lambda_{*}(t;z,x,l,b) &= d\Lambda_{1,z}(t) \exp(\beta_{1,z}' x + \gamma_{1,z} l + \alpha_z b), \\
d\Lambda_{n_1}(t;z,x,l,b) &= d\Lambda_{2,z}(t) \exp(\beta_{2,z}' x + \gamma_{2,z} l + \eta_{2,z}n_1 + \alpha_z b),
\end{align*}
For identifiability issues, we assume the frailty $b \sim N(0,1)$ and the frailty effect $\alpha_z>0$, since a greater value of $b$ may represent an individual being in weaker conditions. The parameters are estimated by NPMLE, and the conditional counterfactual cumulative incidence functions are calculated using the Kolmogorov forward equation.

The computation takes 53 seconds. As in the previous section, we use a bootstrap with 200 resamplings to obtain 95\% confidence intervals. Estimates of coefficients and standard errors are displayed in Panel (B) of Table \ref{tab:est1}. The effect of frailty is estimated at $\hat\alpha_1=0.02$ (95\% CI: 0.01--0.74) in the MUD-HCT ($Z=1$) group and $\hat\alpha_0=0.54$ (95\% CI: 0.13--0.75) in the Haplo-HCT ($Z=0$) group. Relapse increases the hazard of death by about 29 times in the MUD-HCT group (HR: 30.39, 95\% CI: 10.04--91.94) and 5 times in the Haplo-HCT group (HR: 6.02, 95\% CI: 4.32--8.39). GVHD reduces the hazard of relapse by about 16\% in the MUD-HCT group (HR: 0.84, 95\% CI: 0.36--1.99) and 51\% in the Haplo-HCT group (HR: 0.49, 95\% CI: 0.34--0.69), known as the graft-versus-leukemia effect. 

The lower row of Figure \ref{fig:cif1} shows the estimated CIFs of death associated with Haplo-HCT $\hat{F}(t;0)$, MUD-HCT $\hat{F}(t;1)$, and in the hypothetical world $\hat{F}(t;0,1)$ in solid lines. For the interventional effects approach, we display the estimated counterfactual CIFs $\hat{F}(t;0,0)$ and $\hat{F}(t;1,1)$ in dashed lines. The estimated counterfactual CIFs are very close to the potential CIFs. Figure \ref{fig:cif1} also shows the estimated direct and indirect effects. Confidence intervals are calculated using bootstrap resampling. The estimated counterfactual CIFs from the frailty model are similar to those from the model without frailty. The patterns of direct and indirect effects are similar for these two models. We conclude that MUD-HCT reduces the risk of death without affecting the risk of relapse. The time-varying confounder GVHD plays a crucial role in delivering the treatment effect from transplantation to death. In summary, MUD-HCT is associated with a significant survival advantage under the use of PTCy-based GVHD prophylaxis.

Interestingly, the counterfactual CIF of death under the conditional draw is slightly higher than that under the marginal draw. Note that the GVHD risk is lower in the MUD-HCT ($Z=1$) group, so the conditional version draws the relapse process under a lower level of GVHD in the hypothetical world. Due to the weak graft-versus-leukemia effect \cite{kolb2008graft}, the lower risk of GVHD is associated with a higher risk of relapse, and hence higher relapse-related mortality.

In Supplementary Material F, we also consider the separable effects estimand, where we hypothetically assume that GVHD is directly influenced only by the treatment component for relapse. In this way, the hazard of developing GVHD in the hypothetical world would be at the natural level under Haplo-HCT. The estimated direct and indirect effects are similar to those described above. In this example, the interventional effects framework has a more straightforward interpretation than the natural and separable effects frameworks, as we have limited knowledge about the disease progression mechanism following transplantation. The effects of transplant modalities on relapse and death are compound, so dismissible components may not exist for the transplant modality. It is also unclear which component of the treatment induces GVHD. Therefore, the separable effects may be hard to interpret. As a randomized analog to controlled effects, the randomized intervention is counterfactually achievable through a ``by all means'' intervention. The intervention in relapse aligns with clinical interest, as the pre-HCT remission and post-HCT immune management aim to reduce relapse to the lowest level in order to reduce relapse-related mortality. The interventional effects framework does not need to specify which treatment component GVHD is associated with, and it provides a clearer insight into the direct effect: If we can control the risk of relapse, then MUD-HCT leads to lower mortality than Haplo-HCT.

\section{Discussion}

Mediation analysis has been well studied in the context of longitudinal studies. Three frameworks are commonly used to distinguish direct and indirect effects from the total effect: natural, separable, and interventional effects. In the hypothetical world, natural effects control the intermediate event process at the natural level under treatment $z_1$, whereas separable effects manipulate the values of treatment components. In contrast, the interventional effects randomly draw the intermediate event process from the distribution under treatment $z_1$. When no time-varying confounders are present, these three frameworks yield the same statistical functional of treatment effects. In the presence of time-varying confounders, interventional effects require weaker assumptions and fewer restrictions on time-varying confounders. However, the interventional direct and indirect effects sum to a so-called overall effect, which may not be equal to the total effect. Another conditional interventional effects approach draws the intermediate event process conditional on the history of time-varying confounders. The conditional interventional direct and indirect effects sum to the total effect. The randomized intervention shares the spirit of the hypothetical strategy in the ICH E9 (R1) addendum: it may reflect a hypothetical scenario of clinical interest \cite{ICH19}.

As a criticism of the interventional effects framework, the sharp meditational null criterion is not satisfied \cite{miles2023causal, diaz2024non}. The mechanism of treatment effects can be investigated more comprehensively by comparing different mediation analysis frameworks, such as separable effects and conditional interventional effects. Recall that in Figure \ref{fig:graph}, the interventional effects framework aims to distinguish the mixed effects exerted through the path from treatment through the time-varying confounder to the intermediate event and terminal event, which results in a hypothetical intermediate event process following a recanting twin of the time-varying confounder. In contrast, the separable effects framework classifies the effect exerted through this path as the treatment effect on the intermediate event, while the conditional interventional effects framework classifies the effect exerted through this path as the treatment effect on the terminal event. These three frameworks give different interpretations of direct and indirect effects.

We highlight some challenges and outline future study directions. First, the treatment effects in our article are time-dependent. Summarizing the treatment effects into a scalar, like the hazards ratio, odds ratio, and restricted mean survival time lost, is possible. However, a scalar of treatment effect may lose information. Choosing a proper estimand to reflect the overall treatment effect requires respecting clinical interest. Hypothesis tests with $P$-values are also beneficial for summarizing the confidence of the detected treatment effects. Second, apart from the one-dimensional frailty, there can be more complex unmeasured confounding. Auxiliary variables, such as instrumental and negative control variables, may help recover information about unmeasured confounding \cite{Rudolph2024Using, li2025regression}. Third, efficient and multiply-robust estimation methods are appealing. Likelihood-based estimation can be computationally demanding, although it is efficient when all models are correctly specified. Multiply robust estimation can incorporate working models of marginal distributions and the propensity score, which is usually computationally more efficient. However, due to the complexity of time-varying confounders and the distribution of random draws, deriving efficient influence functions and constructing an efficient estimator of treatment effects from working models can be challenging. Fourth, interval censoring is common for time-to-event data. It is interesting to establish identifiability and propose estimators with desired asymptotic properties when the intermediate event is subject to interval censoring.

\section*{Acknowledgments}

This dataset was collected by the Center for International Blood and Marrow Transplant Research (CIBMTR) which is supported primarily by the Public Health Service U24CA076518 from the National Cancer Institute; the National Heart, Lung, and Blood Institute; the National Institute of Allergy and Infectious Diseases; 75R60222C00011 from the Health Resources and Services Administration; N00014-23-1-2057 and N00014-24-1-2507 from the Office of Naval Research; NMDP; and the Medical College of Wisconsin.

\bibliographystyle{apalike}
\bibliography{ref}

\includepdf[pages=1-19]{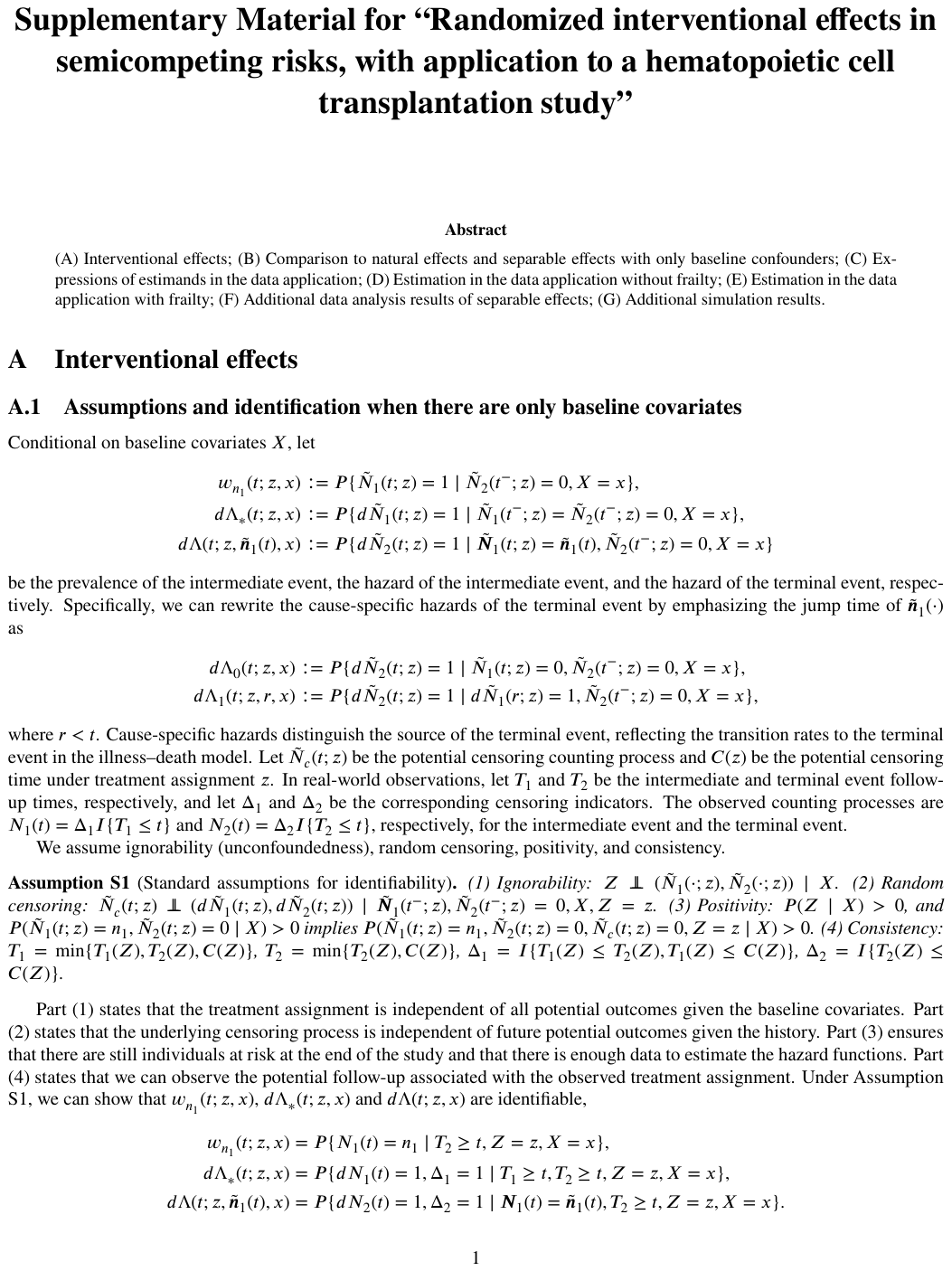}

\end{document}